\documentclass[useAMS,usenatbib,longnamesfirst,usedcolumn]{mn2e} 
\usepackage{epsfig}	
\usepackage{times}	
\usepackage{subfigure}  
\usepackage[a4paper=true,pagebackref,
  pdftitle={LoCuSS: the connection between brightest cluster galaxy activity, gas cooling and dynamical disturbance of X-ray cluster cores}
  pdfauthor={Alastair J.~R. Sanderson},
  pdfsubject={MNRAS},
  pdfkeywords={LoCuSS Chandra BCG cool-core},
  pdfproducer={LaTeX with hyperref},
  pdfcreator={LaTeX}
  pdfdisplaydoctitle=true,
  colorlinks=true,
  citecolor=blue,
  linkcolor=blue,
  urlcolor=blue]{hyperref}

\title[BCG activity and X-ray cluster cores]{LoCuSS: the connection 
between brightest cluster galaxy activity, gas cooling and dynamical 
disturbance of X-ray cluster cores}

\author[A. J.~R. Sanderson, A. C. Edge and G. P. Smith]
       {Alastair J.~R. Sanderson,\thanks{E-mail: ajrs@star.sr.bham.ac.uk}$^1$
        Alastair C. Edge$^2$ and Graham P. Smith$^1$\\
 $^1$School of Physics and Astronomy, University of
        Birmingham, Edgbaston, Birmingham B15 2TT, UK\\
 $^2$Institute for Computational Cosmology, Department of Physics, 
    University of Durham, South Road, Durham DH1 3LE
       \\}
 \date{Accepted 2009 June 8.
      Received 2009 June 8;
      in original form 2009 March 27 ($svn$ $Revision: 77 $)}

\pagerange{\pageref{firstpage}--\pageref{lastpage}}
\pubyear{20??}

\shortcites{best07,bildfell08,cavagnolo08,condon98,crain07,crawford99,
dunkley09,edwards07,katayama03,maughan08,mccarthy08,mittal09,
okabe09,peres98,poole07,san03,smith05,stott08,sun09,vikhlinin07,
vikhlinin06,vikhlinin98b,voit08,zhang08}

\newcommand{\rmsub}[2]{\ensuremath{#1_{\mathrm{#2}}}} 
\newcommand{\Chandra}{\emph{Chandra}}
\newcommand{\chisq}{\ensuremath{\chi^2}}
\newcommand{\fbary}{\rmsub{f}{b}}
\newcommand{\fgas}{\rmsub{f}{gas}}
\newcommand{\Halpha}{H$\alpha$}
\newcommand{\km}{\ensuremath{\mbox{~km}}}
\newcommand{\kmpspMpc}{\ensuremath{\km \ps \pMpc\,}}

\newcommand{\Mpc}{\ensuremath{\mbox{~Mpc}}}
\newcommand{\Msol}{\ensuremath{\mathrm{M}_{\odot}}}

\newcommand{\pMpc}{\ensuremath{\Mpc^{-1}}}
\newcommand{\ps}{\ensuremath{\s^{-1}}}
\newcommand{\rfiveh}{\rmsub{r}{500}}
\newcommand{\rtwofiveh}{\rmsub{r}{2500}}
\newcommand{\s}{\ensuremath{\mbox{~s}}}

\newcommand{\Yx}{\rmsub{Y}{X}}

\begin{document}

\maketitle

\label{firstpage}

\begin{abstract} 
 \noindent We study the distribution of projected offsets between the
 cluster X-ray centroid and the brightest cluster galaxy (BCG) for 65 X-ray
 selected clusters from the Local Cluster Substructure Survey (LoCuSS),
 with a median redshift of $z$\,=\,0.23. We find a clear correlation
 between X-ray/BCG projected offset and the logarithmic slope of the
 cluster gas density profile at 0.04\rfiveh\ ($\alpha$), implying that more
 dynamically disturbed clusters have weaker cool cores. Furthermore, there
 is a close correspondence between the activity of the BCG, in terms of
 detected \Halpha\ and radio emission, and the X-ray/BCG offset, with the
 line emitting galaxies all residing in clusters with X-ray/BCG offsets of
 $\le$15\,kpc. Of the BCGs with $\alpha<$\,$-$0.85 and an offset
 $<$\,0.02\rfiveh, 96 per cent (23/24) have optical emission and 88 per
 cent (21/24) are radio active, while none has optical emission outside
 these criteria. We also study the cluster gas fraction (\fgas) within
 \rfiveh\ and find a significant correlation with X-ray/BCG projected
 offset. The mean \fgas\ of the `small offset' clusters ($<$\,0.02\rfiveh)
 is $0.106\pm0.005$ ($\sigma=0.03$) compared to $0.145\pm0.009$
 ($\sigma=0.04$) for those with an offset $>$\,0.02\rfiveh, indicating that
 the total mass may be systematically underestimated in clusters with
 larger X-ray/BCG offsets. Our results imply a link between cool core
 strength and cluster dynamical state consistent with the view that cluster
 mergers can significantly perturb cool cores, and set new constraints on
 models of the evolution of the intracluster medium.
\end{abstract}

\begin{keywords}
  galaxies: clusters: general -- X-rays: galaxies clusters -- cooling flows
  -- galaxies: evolution -- galaxies: elliptical and lenticular, cD

\end{keywords}

\section{Introduction}
The first ranked, or brightest cluster galaxies (BCGs) are amongst the
brightest of all galaxies, accounting for around 5--10 per cent of the
total light in massive clusters \citep{lin04b}. Their properties are
closely related to those of the host cluster \citep{edge91a,lin04b} and
they typically lie at the bottom of the potential well. In relaxed
clusters, this location is frequently positioned within a `cool core' in
the intracluster medium (ICM), where gas is capable of condensing out of
the hot phase, to form stars \citep[e.g.][]{crawford99,rafferty08} or fuel accretion
onto a central supermassive black hole \citep[e.g.][]{best07}. In this
configuration the BCG lies at a interface which is crucial to understanding
the role of feedback in galaxy and cluster evolution.

The unique character of BCGs allows them to be used as important
diagnostics of the host cluster. In particular, the projected offset
between the X-ray peak and the BCG is sensitive to the cluster dynamical
state \citep{katayama03}, which is analagous to the offset between the
peaks of the gravitational lensing mass map and the X-ray emission
as an indicator of disturbance \citep{smith05}. Furthermore, the presence
of \Halpha\ emission from ongoing star formation
\citep[e.g.][]{heckman81,peres98,crawford99,edwards07,cavagnolo08} and radio 
emission from active galactic nuclei (AGN) activity
\citep[e.g.][]{peres98,best07,cavagnolo08,mittal09} probe the thermodynamic state
of the hot gas in the cluster core. For example, it has recently been
discovered that star formation in BCGs only occurs when the entropy of the
ICM falls below a critical threshold
\citep{voit08,rafferty08}. Moreover, almost all clusters with star-forming
BCGs lie above the X-ray luminosity-temperature relation
\citep{bildfell08}, demonstrating the close link between the BCG and global
cluster properties.

With the advent of \Chandra\ it is possible to probe cluster gas properties
on the scale of BCGs out to intermediate redshifts and with its extensive
archive of observations, this type of analysis can be applied to large
numbers of clusters. In this paper we use \Chandra\ data to explore the
connection between cooling in cluster cores and the location and activity
of the BCG for an X-ray selected sample of 65 clusters drawn from the Local
Cluster Substructure Survey\footnote{http://www.sr.bham.ac.uk/locuss}
(LoCuSS). This is a morphologically unbiased sample of $\sim$100 X-ray
luminous galaxy clusters selected from the (e)BCS and REFLEX ROSAT
All Sky Survey (RASS) catalogues \citep[e.g. see][]{zhang08,okabe09},
containing all clusters down to the RASS flux limit, within the LoCuSS
redshift, declination and Galactic column cuts. Throughout this paper we
assume $H_{0}=70$\kmpspMpc, $\rmsub{\Omega}{m}=0.3$ and
$\Omega_{\Lambda}=0.7$.  Errors are quoted at a 68 per cent confidence
level.

\section{sample selection and data analysis}
We select all clusters with available Chandra data that satisfy the LoCuSS
selection function of $0.15\le z\le0.3$, $-70^\circ\le\delta\le70^\circ$,
$n_{\mathrm{H}}<7\times10^{20}\,{\rm cm}^{-2}$. This produces a total of 66
clusters, from which we discard ZwCl\,1309.1$+$2216 because the data on
this cluster are too shallow for our purposes. The median redshift is
$z$\,=\,0.23 with a corresponding scale factor of 3.7\,kpc per arcsecond.
Eighteen of the 65 cluster observations are drawn from our own Chandra
programs in Cycles 9 and 10 (PIDs: 09800732, 10800565), the remaining
observations are from the archive.

\subsection{X-ray data analysis}
\Chandra\ data were reduced and analysed according to the procedure described
in \citet{san09}. Briefly, the \Chandra\ data were reprocessed using
\textsc{ciao} version 3.4 and incorporating \textsc{caldb} version 3.4.2, 
to produce flare cleaned and point source removed level 2 events
files. Corresponding blank sky background datasets were also produced and
matched in normalization to the cluster events in each case, to account for
variations in the particle-dominated high energy background. No adjustment
was made to the background to allow for any variation in soft Galactic
foreground emission compared to each cluster observation. However,
following \citet{san06}, the galactic absorbing column was fitted as a free
parameter in the spectral modelling to allow for any differences in
inferred low energy absorption associated with soft emission excesses or
calibration uncertainties. For some clusters, it was necessary to freeze
the absorbing column at the galactic HI value, since unfeasibly low values
were otherwise obtained in many of the annular bins.

A series of annular spectra were extracted for each cluster, comprising
between 6 and 25 radial bins in total, depending on data quality. These
annular spectra were then fitted with an absorbed \textsc{apec} model using
the \textsc{projct} scheme in \textsc{xspec} version 11.3.2 to yield
deprojected gas temperature and density profiles for each cluster. Weighted
response matrix files were used for each spectrum, and the fitting was
performed in the energy range 0.5--7.0\,keV. In order to provide more
reliable error estimates on the annular spectral measurements, we performed
200 Poisson realizations of the best-fit \textsc{projct} model, using the
same background and response files (Sanderson, in preparation). Each
realization was treated like the original data and fitted with the
\textsc{projct} scheme to produce a suite of simulated measurements in 
each bin. The errors in each bin were then calculated from the median
absolute deviation of these simulated values, as a robust estimator of the
standard deviation that is well-suited to heavy-tailed distributions
\citep[see][for example]{beers90}.

We note that a potential issue with our \Chandra\ analysis is the
possibility of a modest bias in temperature estimates for hotter clusters
($\ga$4-5 keV) resulting from errors in the \Chandra\ response matrix
\citep[as described in][and references therein]{sun09}, which could lead to
overestimates of the temperature. To assess the magnitude of the effect
this could have on our analysis, we have reanalysed one of the cluster
datasets using \textsc{ciao} version 4.1, incorporating
\textsc{caldb} version 4.1.2. We have selected Abell~1835 (obsid 6880), 
since this is a deep observation (120\,ks) of a bright, massive cluster,
where the impact of calibration changes is likely to be greatest. Using the
new calibration data, we find that \rfiveh\ changes to $1432\pm60$\,kpc
(from $1506\pm57$, in our original analysis), with a corresponding gas
fraction within \rfiveh\ of $0.128\pm0.011$ ($0.110\pm0.008$
originally). As expected, the newer \textsc{caldb} results in a lower
\rfiveh\ and hence higher gas fraction (due to both a systematic 
lowering of the temperature leading to a lower total mass, as well as an
increase in the gas mass measured from the emissivity), but the effect is
only $\sim$5 per cent for \rfiveh\ and 16 per cent for the gas fraction;
corresponding to 1.2 and 1.6$\sigma$, respectively. Since only two of our
clusters are cooler than 4\,keV, such a systematic shift will be similar
across the whole sample and so is not likely to impact our results
significantly.

The centroid for the spectral profile analysis was used to determine the
projected offset between the centre of the cluster X-ray halo and the
BCG. We determined this position in a similar fashion to \citet{maughan08},
iteratively refining the cluster X-ray surface brightness centroid within
an aperture of 2--3 arcmin initially, then repeating the centroiding within
an aperture of $\sim$1 arcmin. In addition, we also determined the
coordinates of the X-ray peak as an alternative reference position to
compare with the BCG location (see Section~\ref{ssec:offset}). To minimize
the impact of Poisson noise, we constructed a smoothed 0.5--2.0\,keV image
for each cluster using the wavelet decomposition method of
\citet{vikhlinin98b}. The X-ray peak was determined as the highest pixel
nearest to the X-ray centroid, ignoring any non-cluster point sources,
which were identified by being detected on only the smallest 1--2 wavelet
scales. We have neglected the (small) uncertainty in the measurement of the
X-ray centroid and peak, and therefore also in the corresponding projected
distance to the BCG, since this is insignificant compared to the
uncertainties on other parameters.

\subsection{BCG identification and properties}
The positions of the BCG optical peaks were obtained from several sources:
near-infrared imaging from the Hale 200-in telescope at Palomar
Observatory\footnote{The Hale Telescope at Palomar Observatory is owned and
operated by the California Institute of Technology.}, described in detail
in \citet{stott08}; near-infrared data from the CTIO 4-m
telescope\footnote{Operated by the Association of Universities for Research
in Astronomy Inc.} (Hamilton-Morris et al., in preparation); optical
imaging from VLT/FORS2\footnote{Operated by the European Southern
Observatory} (PI Edge), as well as the Digitized Sky Survey. Photometry of
these imaging data were supplemented by spectroscopic confirmation of
membership, using spectra from several sources (see below). The BCG was
identified as the brightest galaxy in the central few hundred kpc of the
cluster. In a few cases the BCG could not be identified unambiguously, but
these appear to be merging clusters in which, regardless of the
ambiguity, the X-ray/BCG projected offset is large.

\begin{figure*}
  \centering
  \subfigure{
    \includegraphics[width=11cm]{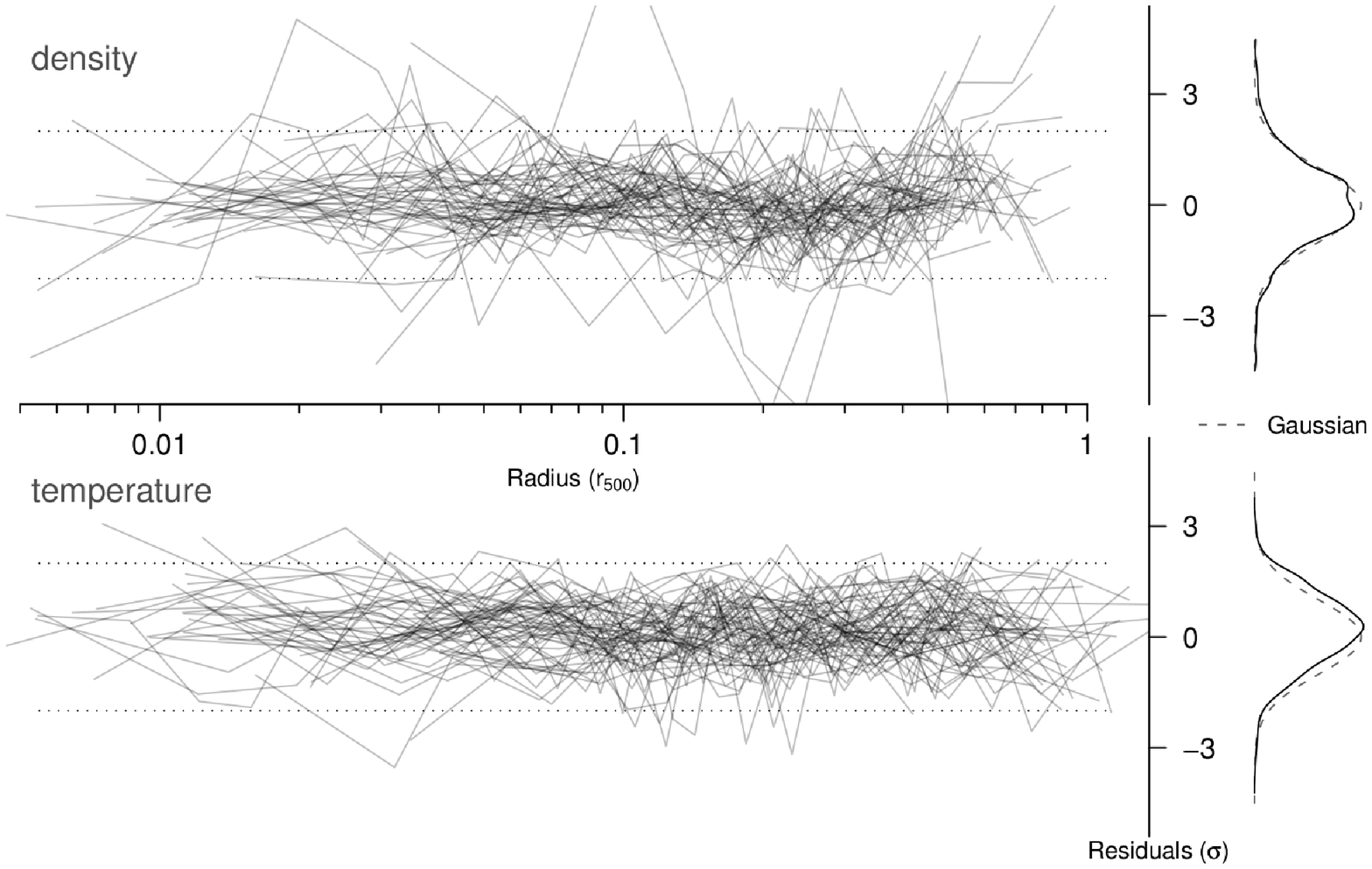}}
  \hspace{0cm}
  \subfigure{
    \includegraphics[width=6.3cm]{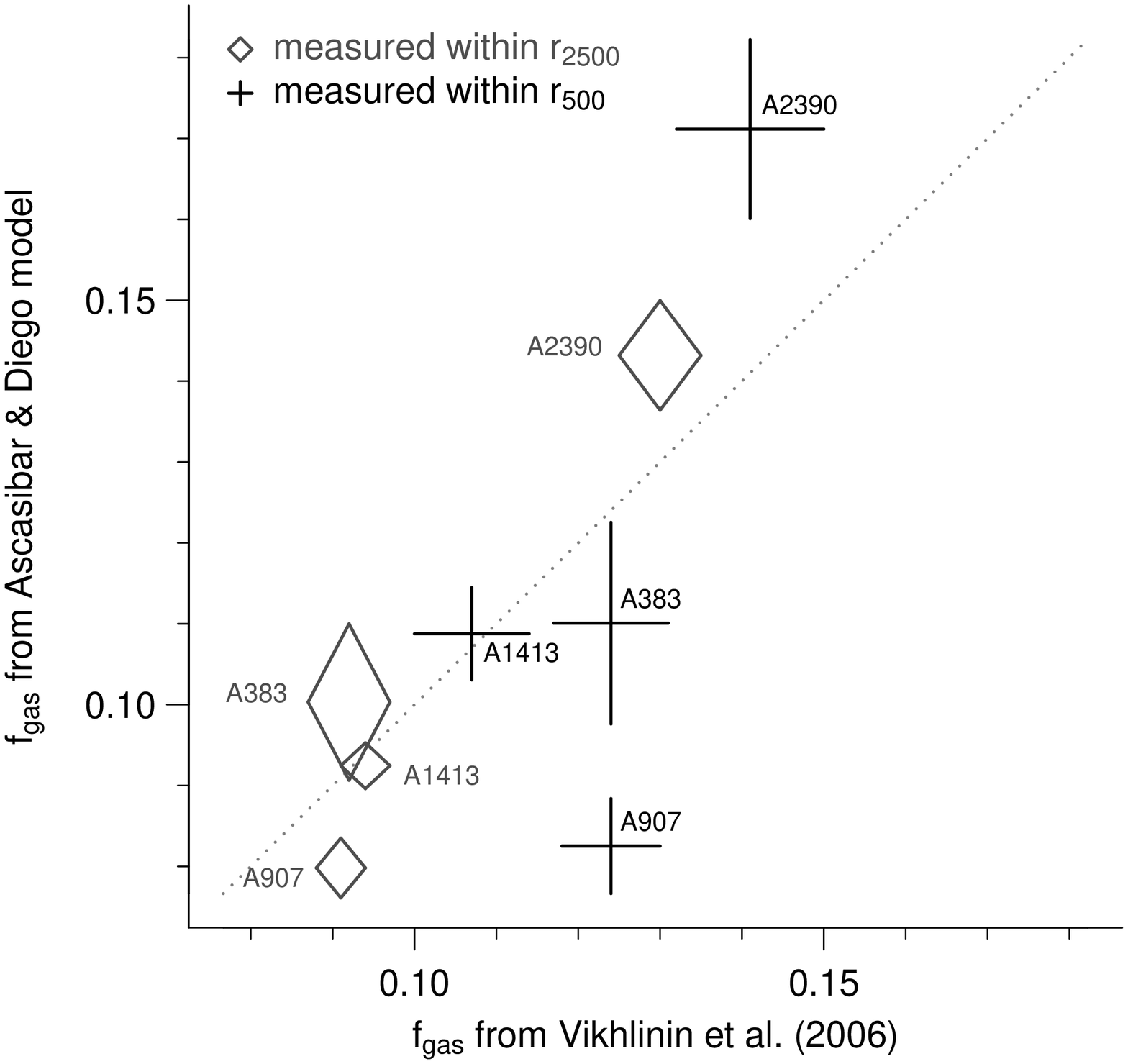}}
  \caption{\textit{Left:} A comparison of the residuals from the 
   \citet{ascasibar08} model, normalized by the measurement errors on each 
   point, as a function of scaled radius for both the gas density and 
   temperature data; each line represents a different cluster. Dotted lines 
   mark $\pm2\sigma$ and the marginal distribution of the data is depicted 
   as a kernel smoothed density estimate, compared to a Gaussian of unit 
   variance. \textit{Right:} A comparison of gas fractions measured within
   both \rfiveh\ and \rtwofiveh\ for the four clusters in common with 
   the sample of \citet{vikhlinin06}; the dotted line marks the locus of 
   equality.}
  \label{fig:resid}
\end{figure*}

Each galaxy was classified according to the presence or absence of \Halpha\
emission lines, as determined from the literature \citet{crawford99}, Sloan
Digital Sky Survey archival spectra \citep{adelman-mccarthy08} and VLT
FORS2 spectra for REFLEX BCGs (PI Edge). For two clusters (Abell 115b and
RXC J2211.7-0350) there was no available optical spectrum to determine the
emission line status of the BCG. \Halpha\ emission is a reliable indicator
of the presence of central cold material, second only to direct detection
of molecular gas at sub-mm wavelengths \citep{edge01}. In addition, we have
classified each BCG according to the presence of radio emission detected in
the Very Large Array (VLA) FIRST Survey \citep{becker95}, NVSS
\citep{condon98} and/or SUMSS \citep{bock99} radio surveys. While these
three radio surveys have different flux density limits, resolution and
frequency, they nevertheless provide a uniform limit of 3\,mJy at 1.4\,GHz.

\section{X-ray cluster modelling}
The deprojected gas temperature and density profiles were fitted jointly
to the phenomenological cluster model of \citet{ascasibar08}, using the
\chisq\ statistic, in order  to determine the gravitating mass profile and 
parametrize the gas thermodynamic structure (Sanderson, in preparation).
The best-fit cluster model was also used to estimate \rfiveh, the radius
enclosing a mean overdensity of 500 with respect to the critical density of
the Universe at the cluster redshift.  Any spectral bins lying within
5\,kpc of the centre were excluded from the fit, owing to systematic
deviations from the model which are expected to be significant on small
scales \citep{ascasibar08}. For the clusters ZwCl~0839.9+2937 and
Abell~1758N it was necessary to exclude the central bins from the fit
(lying at 10 and 30\,kpc, respectively) due to significant deviations from
the best-fit model.

The left panel of Fig.~\ref{fig:resid} shows the residuals from the model
for both the gas temperature and density, as a function of scaled
radius. It is clear that the model provides a good description of the data,
with no indication of any significant systematic deviations. The marginal
distribution of the residuals is consistent with the Gaussian distribution
expected from the measurement errors alone, with only a small shift
discernable in the case of the temperature profile-- indicating that on
average the model tends to slightly ($\sim$0.3$\sigma$) underpredict the
data. There is particularly good agreement between the data and the model
in the gas density at all radii, including 0.04\rfiveh, where we use the
logarithmic gradient to quantify the strength of cooling
(Section~\ref{sec:CC}). It should be noted that the gas density profiles
extend only to the penultimate spectral bin, owing to the non-trivial
volume element associated with the outermost annulus in the deprojection
\citep{san06}.

As an additional test of the model, the right panel of Fig.~\ref{fig:resid}
plots the derived gas fraction (\fgas) within both \rtwofiveh\ and \rfiveh\
compared to measurements made by \citet{vikhlinin06}, for the four clusters
common to both samples. We calculate errors on \fgas\ and all other derived
quantities directly, using the median absolute deviation of 200 bootstrap
resamplings of the input temperature and density profile data. There is
generally reasonable agreement in \fgas, particularly for the measurements
within \rtwofiveh, and no indication of serious systematic trends, although
both Abell~907 and Abell~2390 are significant outliers, within
\rfiveh. This disagreement points to the importance of systematics
associated with cluster modelling, for example differences in the number
and location of radial bins, as well as the choice of mass profile
parametrization. Nevertheless, we emphasize the close match between our
cluster model fits and the data, as seen in residuals from the
model. Furthermore, in Fig.~\ref{fig:compare_r500}
(Section~\ref{ssec:dyn_state}) we demonstrate good agreement between
\rfiveh\ compared to measurements made by \citet{maughan08}, where we 
address the issue of bias in X-ray mass estimates in relation to cluster
dynamical state.

\section{Results}
\subsection{X-ray/BCG offset}
\label{ssec:offset}
A number of recent studies have identified the projected separation between
the BCG and the \emph{peak} of the cluster X-ray emission as in important
parameter in understanding the activity of the BCG, both in terms of star
formation \citep[e.g.][]{edwards07,bildfell08,cavagnolo08} and AGN radio
emission \citep[e.g.][]{cavagnolo08,mittal09}. However, while the X-ray
peak can accurately identify the focus of a cool core, this location may
not always lie at the `centre' of the hot gas, as defined for the purposes
of conducting a radial profile analysis. This is particularly true for
non-CC clusters, which have much broader and flatter X-ray cores. We
therefore investigate the projected offset between the BCG and both the
peak \emph{and} centroid of the X-ray emission. We assess the relative
merits of both definitions in the context of quantifying cluster dynamical
state in Section~\ref{sec:CC}, but turn our attention initially to the
projected offset in terms of the centroid.

\begin{figure}
\includegraphics[width=8.4cm]{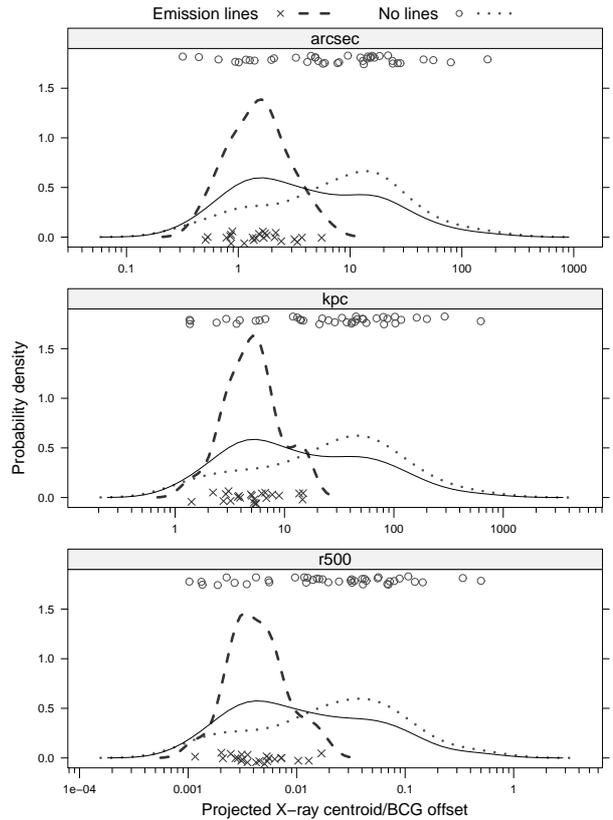}
\caption{ \label{fig:densityplot-Halpha}
Kernel-smoothed probability density estimates of the distribution of
projected offsets between the X-ray cluster centroid and the BCG. The raw
values are shown at randomly `jittered' $y$-axis positions. The
distributions for BCGs with and without detected \Halpha\ emission are also
indicated.}
\end{figure}

Fig.~\ref{fig:densityplot-Halpha} shows the distribution of X-ray
centroid/BCG projected offsets for the sample, in units of arseconds,
kiloparsecs and \rfiveh. The distributions are roughly lognormal, with
medians of 13\,kpc / 1.0 per cent of \rfiveh.  Also plotted are the
separate distributions for BCGs with and without detected \Halpha\ line
emission, and a clear difference is apparent: the BCGs with line emission
all have small projected offsets from the X-ray cluster centroid, within a
maximum of 15 kpc / 1.5 per cent of \rfiveh. Conversely, BCGs with no
detected \Halpha\ emission lines have preferentially larger X-ray
centroid/BCG projected offsets (median values = 35\,kpc / 2.8 per cent of
\rfiveh). This confirms that close proximity of the BCG to the X-ray
cluster centroid is a pre-requisite for BCG star formation, in agreement
with recent work \citep{edwards07,bildfell08,rafferty08}.

\begin{figure}
\centering
\includegraphics[width=8.4cm]{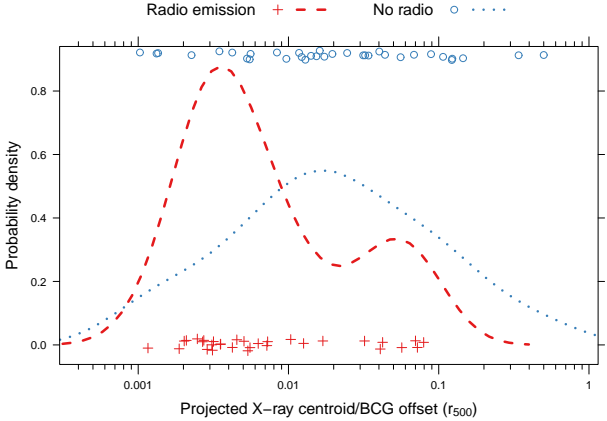}
\caption{ \label{fig:densityplot-radio}
Kernel-smoothed probability density estimates of the distribution of
projected X-ray/BCG offsets, normalized to \rfiveh. The data are separated
according to whether or not radio emission is detected from the BCG and the
raw values are shown at randomly `jittered' $y$-axis positions.}
\end{figure}

For comparison, Fig.~\ref{fig:densityplot-radio} shows the distribution of
projected offsets for BCGs with and without detected radio emission. As with
the \Halpha\ emitters in Fig.~\ref{fig:densityplot-Halpha}, the clusters
with radio detected BCGs have smaller X-ray/BCG offsets. However, a
separate peak of radio detected BCGs with intermediate projected offsets is
also present. For the sample as a whole, the interquartile range of
X-ray/BCG projected offsets is 0.0035--0.04\rfiveh\ (see bottom right panel of
Fig.~\ref{fig:alpha_vs_offset}), demonstrating that BCGs in X-ray selected
clusters are generally located close to the cluster X-ray centroid; all but
two (97$^{+2}_{-4}$ per cent) lie within 0.15\rfiveh.

\subsection{BCG \Halpha\ and radio emission}
\label{ssec:BCG_activity}
Table~\ref{tab:Halpha-radio} summarizes the \Halpha\ and radio emission
status of the 63 clusters for which both classifications are available.
Roughly half the galaxies are passive (30/63), with no detected \Halpha\ or
radio emission. The radio active fraction is 49$\pm7$ per cent, which is
larger than the value of $\ga$\,30 per cent found in clusters selected
optically \citep{best07} and in X-rays \citep{lin07}, possibly due to the
high average mass of the clusters in the LoCuSS sample. Similarly, the
fraction of \Halpha\ emitting BCGs ($37\pm7$ per cent) is somewhat higher
than the value of 27 percent for the X-ray selected cluster sample of
\citet{crawford99}, which is itself roughly double the fraction in
optically-selected clusters \citep[13 per cent;][]{edwards07}. A Pearson's
\chisq\ test strongly rejects the null hypothesis that the presence or
absence of detected \Halpha\ emission is independent from the presence or
absence of radio emission, with a \chisq\ value of 25.7 for 1 degree of
freedom and a corresponding $p$-value of $4.0\times10^{-7}$. This
demonstrates the close correspondence between BCG activity in terms of star
formation and black hole accretion.

\begin{table}
\begin{center}
\begin{tabular}{rrr|l}
               & Emission lines & No lines & Sum \\
\\[-2ex]
Radio emission &           21 (0.33)  &   10 (0.16)    &  31 (0.51) \\
      No radio &            2 (0.03)  &   30 (0.48)    &  32 (0.49) \\
\\[-2ex]
\hline \\[-2ex]
Sum &           23 (0.36)  &   40 (0.64)    &  63 (1) \\
\end{tabular}
\caption{A contingency table of the BCG distribution according to their 
\Halpha\ and radio emission status, with fractions of the total 
(63) given in brackets. Note that the two clusters with unknown 
emission line status are excluded.}
\label{tab:Halpha-radio}
\end{center} 
\end{table}

\subsection{Cluster cool core status}
\label{sec:CC}
In order to connect the properties of the BCG to those of its host cluster,
we focus on the cooling state of the intracluster medium (ICM). A key
signature of cool core clusters is a cuspy gas density-- and hence surface
brightness-- profile, which can be more accurately measured than the
associated decline in temperature. Furthermore, the gradient of the gas
density profile on small scales progressively steepens with increased
cooling, even before a significant cool core is established
\citep{ettori08}. Therefore, following \citet{vikhlinin07}, we use the
parameter $\alpha$, defined as the logarithmic slope of the gas density
profile at 0.04\rfiveh, to quantify the extent of cluster gas cooling on
the approximate scale of any central galaxy, with more negative values
implying stronger cooling.

\begin{figure}
\centering
\includegraphics[width=8.4cm]{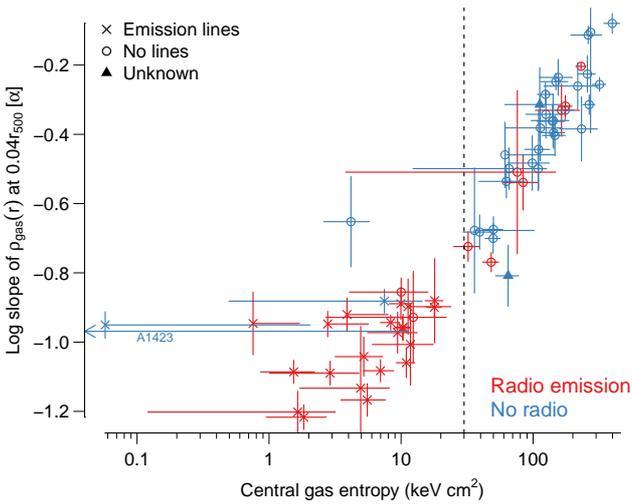}
\caption{ \label{fig:alpha_vs_egas0}
$\alpha$ (see Section~\ref{sec:CC}) versus central gas entropy, from the
best-fit cluster model. The vertical dashed line marks
30\,keV\,cm$^2$. Point styles indicate the presence or absence of \Halpha\
emission lines in the BCG and the colour shows whether significant radio
emission is detected from the BCG. The zero value for Abell~1423 is plotted
as an arrow from its upper bound.}
\end{figure}

We calculate the logarithmic slope, $\alpha$, from the best-fit
\citet{ascasibar08} cluster model and plot it, in
Fig.~\ref{fig:alpha_vs_egas0}, against the central gas entropy, which is a
direct indicator of the cooling state of the ICM.  However, this
measurement involves extrapolation of the model to zero radius, unlike
$\alpha$, which is evaluated at a directly accessible radius, which also
lies in a region where the model provides a good fit to the density profile
(Fig.~\ref{fig:resid}, left panel). The strong correlation in
Fig.~\ref{fig:alpha_vs_egas0} demonstrates the effectiveness of $\alpha$ as
a simple means of quantifying cool-core strength.

It can be seen from Fig.~\ref{fig:alpha_vs_egas0} that there is a gap in
the entropy values which separates the clusters into two regimes, with all
the \Halpha\ emitting BCGs lying in clusters with central entropy
$<$30\,keV\,cm$^2$, as discovered by \citet{rafferty08} and subsequently
confirmed by \citet{cavagnolo08}. Furthermore, there are a number of
clusters with a central gas entropy of $\sim$10\,keV\,cm$^2$, with the
remainder grouped around $\sim$100--200\,keV\,cm$^2$ as also found by
\citeauthor{cavagnolo08}. This threshold entropy level can be understood as
being determined by the physics of thermal conduction \citep{voit08}, which
can also account for the bifurcation into CC and non-CC populations
\citep[e.g.][]{guo08,san09}.

\begin{figure*}
  \centering
  \subfigure{
    \includegraphics[width=8.7cm]{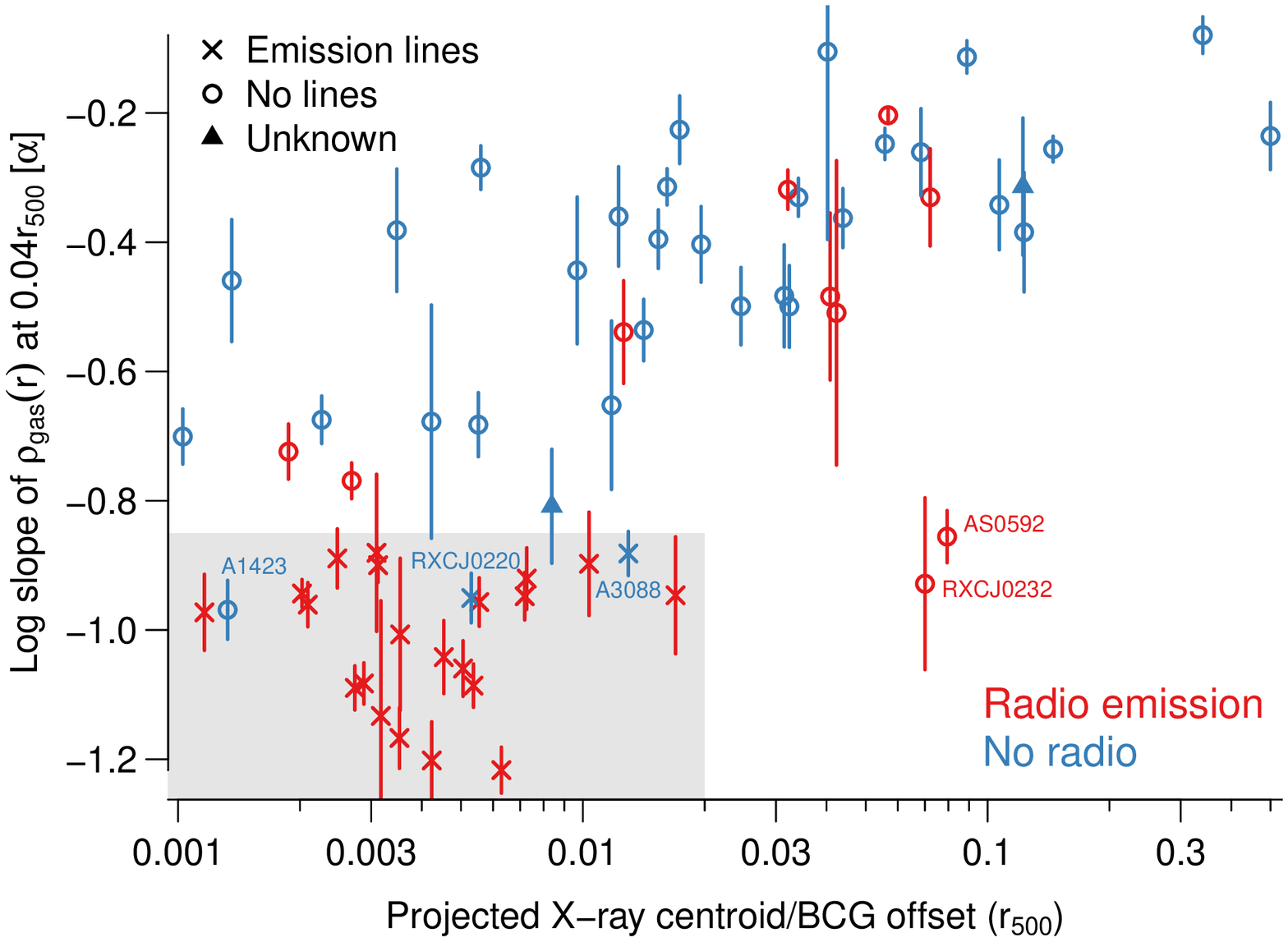}}
  \hspace{0cm}
  \subfigure{
    \includegraphics[width=8.7cm]{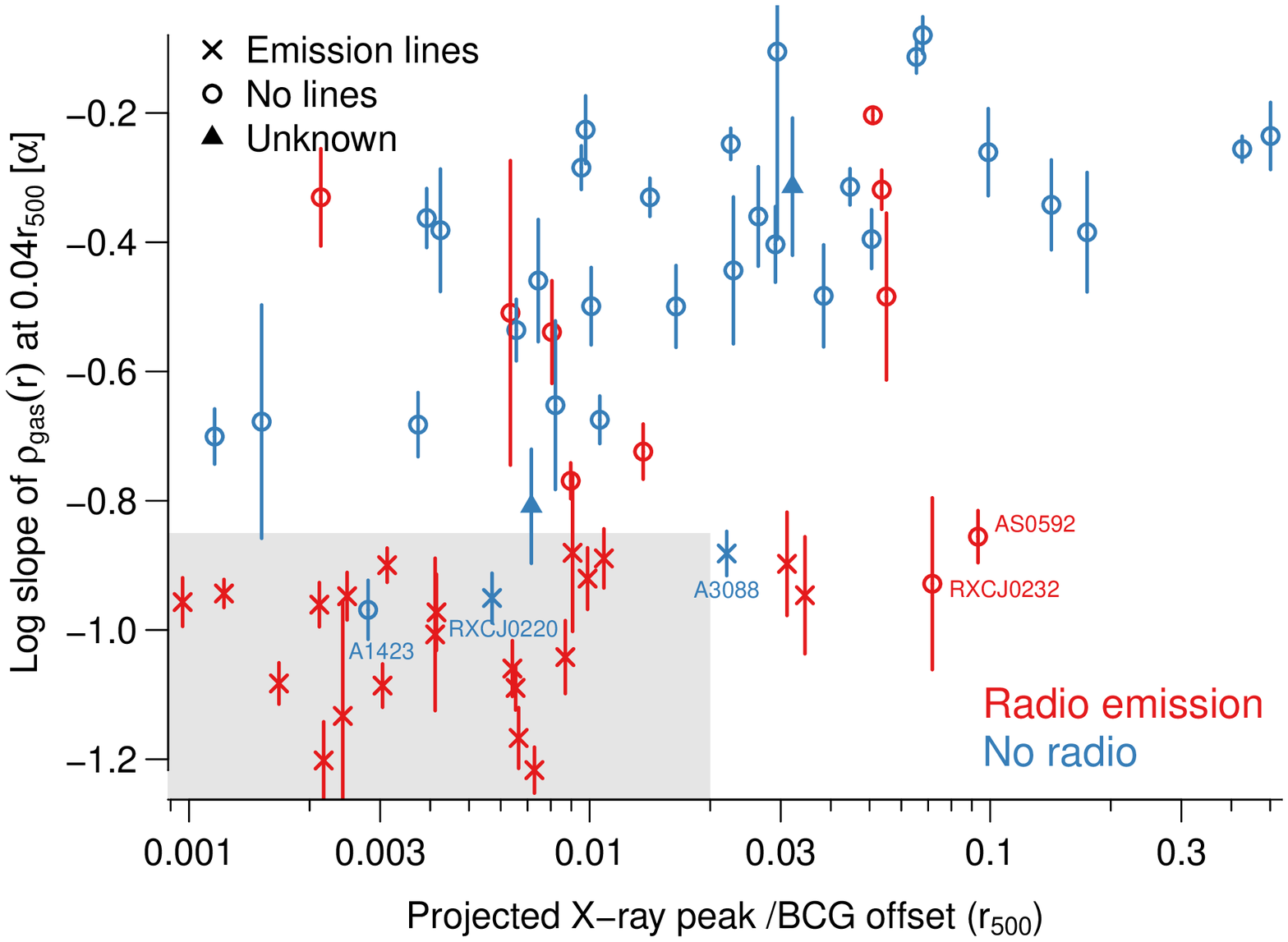}}
  \caption{$\alpha$ (see Section~\ref{sec:CC}) plotted
  against the projected offset between the BCG and the X-ray \emph{centroid} 
  (left panel) and the X-ray \emph{peak} (right panel). The shaded box 
  highlights the region occupied by line emitting galaxies in the left panel.}
  \label{fig:alpha_vs_offset}
\end{figure*}

Given the greater ease and reliability with which $\alpha$ can in general
be measured, we use this as the parameter of choice for quantifying cool
core strength. We now turn to the issue of how cluster cooling is related
to the X-ray/BCG offset. Fig.~\ref{fig:alpha_vs_offset} shows the variation
of $\alpha$ with X-ray/BCG projected offset (scaled by \rfiveh) defined
using both the X-ray centroid (left panel) and peak (right panel). A clear
trend is evident, with the strongest cooling cores found only in clusters
with small projected X-ray/BCG offsets. \citet{vikhlinin07} suggest
$\alpha<$\,$-0.7$ for strong cooling flows, which lies close to the gap in
the distribution of values. However, it is clear that only those clusters
with $\alpha\la-0.85$ \emph{and} a projected offset $\la0.02$\rfiveh\ have
detected \Halpha\ line emission (as depicted by the shaded box), which is
indicative of genuine gas condensation and subsequent star formation
associated with unchecked cooling. Of the 24 systems in this shaded region,
only 1 is passive, and 21 have both \Halpha\ and radio emission; there are
no \Halpha\ emitting BCGs outside the shaded region.

While the two panels in Fig.~\ref{fig:alpha_vs_offset} show similar
results, the correlation between $\alpha$ and the X-ray \emph{centroid}/BCG
projected offset is stronger and shows fewer outliers: the Kendall rank
correlation coefficient gives $\tau$ = 0.48 ($p$-value =
$1.3\times10^{-8}$) for the offset with respect to the centroid and
$\tau$ = 0.42 ($p$-value = $6.7\times10^{-7}$) for the peak. Nevertheless,
a number of prominent outliers are still present in the left panel of
Fig.~\ref{fig:alpha_vs_offset} and are labelled in both panels. In
particular Abell~S0592 and RXC~J0232.2-4420, which have relatively large
X-ray/BCG offsets, despite hosting strong cool cores. Abell~S0592 has an
obvious subclump (excluded from the X-ray analysis) which is associated
with the BCG; the main peak X-ray centroid is located close ($<$10\,kpc) to
another galaxy. RXC~J0232.2-4420, on the other hand, has a relatively
regular and single-peaked morphology, but with 2 bright galaxies of which
the less luminous is located close to ($<$10\,kpc) the X-ray centroid.

The only passive cluster with a steep $\alpha$ and small projected
X-ray/BCG offset is Abell~1423.  While there is a extended radio source
very close to the BCG, inspection of the FIRST Survey image indicates that
this is an unrelated narrow angle tail radio galaxy seen in projection and
that the BCG is undetected to the FIRST depth.  No optical lines were
detected by \citet{crawford99} but the spectrum was not of the highest
quality so more detailed follow-up of this BCG would be of great interest.
The two remaining labelled points are the clusters RXC~J0220.9-3829 and
Abell~3088, which are the only other systems in the shaded box with BCGs
that have no significant radio emission.

While it is clear that \Halpha\ emitting BCGs are all concentrated in the
bottom left corner of Fig.~\ref{fig:alpha_vs_offset} (particularly the left
panel), the distribution of passive BCGs (blue circles, with no \Halpha\ or
radio emission) follows a continuous strong trend: a test with the Kendall
rank correlation coefficient for the centroid offset case gives a value of
$\tau$ 0.53 ($p$-value = 2.0$\times10^{-5}$) in favour of the null
hypothesis that the two quantities are uncorrelated.  The equivalent
Kendall coefficient for the peak offset case is $\tau$ 0.48 ($p$-value =
1.0$\times10^{-3}$). This suggests a progressive weakening of central gas
cooling with increasing cluster disruption, as measured by the X-ray/BCG
projected offset, even for clusters with no evidence of condensation from
the hot phase.

\subsection{Cluster gas fraction and dynamical state}
\label{ssec:dyn_state}
\begin{figure*}
  \centering \subfigure{
  \includegraphics[width=8.7cm]{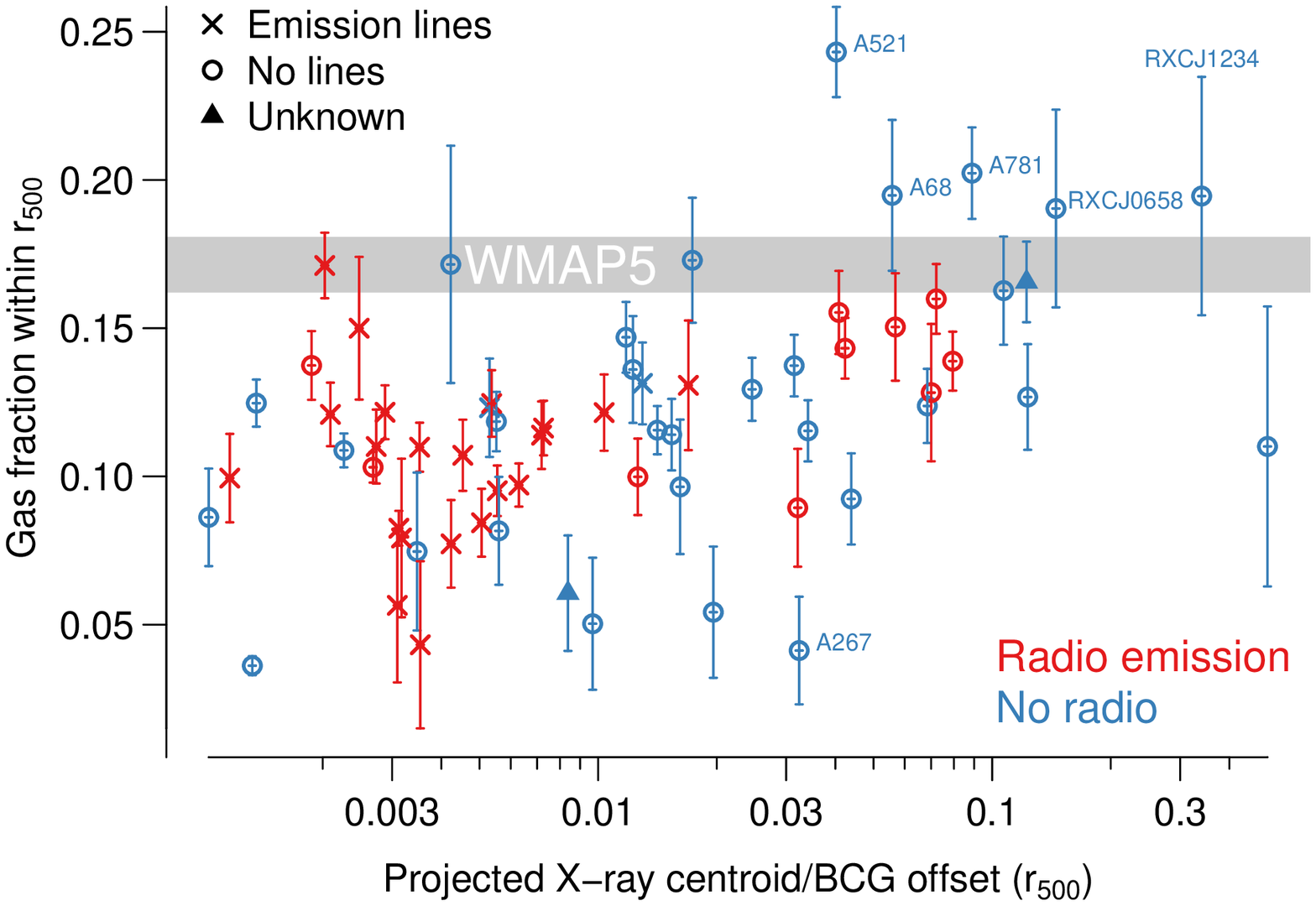}}
  \hspace{0cm} \subfigure{
  \includegraphics[width=8.7cm]{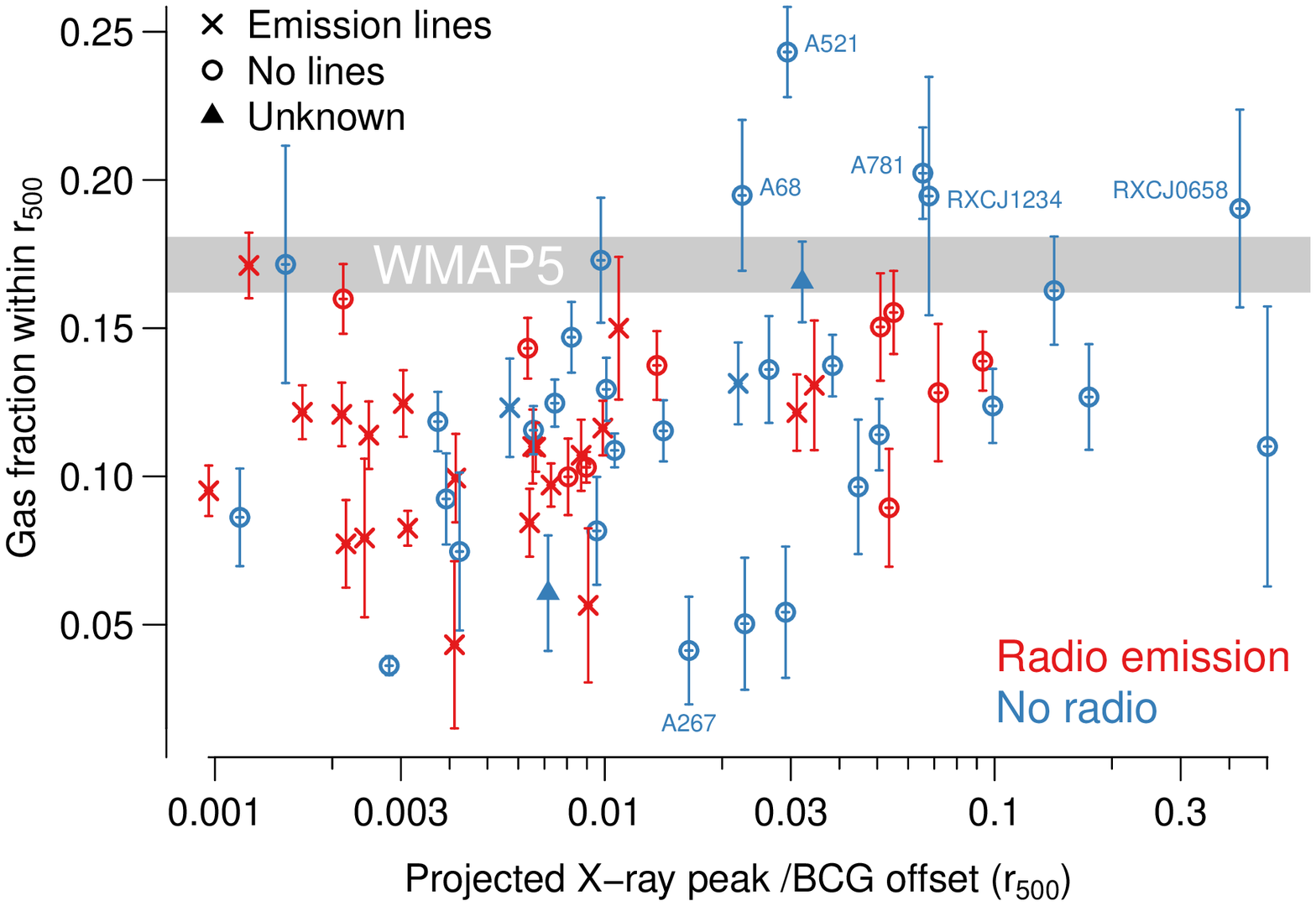}}
  \caption{Cluster gas fraction within \rfiveh\ as a function of projected
  offset between the BCG and the X-ray \emph{centroid} (left panel) and the
  X-ray \emph{peak} (right panel). The horizontal shaded region depicts the
  Universal baryon fraction as measured from the five year WMAP data
  \citep{dunkley09}. Outliers identified in Fig.~\ref{fig:compare_r500} are
  labelled: see text for details.}
  \label{fig:fgas_vs_offset}
\end{figure*}

To explore the connection between X-ray/BCG offset and cluster dynamical
state, we plot in Fig.~\ref{fig:fgas_vs_offset} the gas fraction (\fgas)
within \rfiveh\ as a function of projected X-ray/BCG offset, with the
Universal baryon fraction (\fbary) from the five year WMAP data
\citep{dunkley09} also indicated. As in Fig.~\ref{fig:alpha_vs_offset}, 
the relation is shown for the offset in terms of both the X-ray centroid
(left panel) and peak (right panel). A test with the Kendall coefficient
for the centroid offset case gives $\tau$ = 0.31 ($p$-value =
$3.0\times10^{-4}$) in favour of the null hypothesis that \fgas\ and the
X-ray/BCG offset are uncorrelated. The equivalent Kendall coefficient for
the peak offset case is $\tau$ = 0.24 ($p$-value = 5.0$\times10^{-3}$). In
contrast, there is no significant correlation between the X-ray/BCG offset
and the cluster mean temperature or total mass, which might otherwise
account for such a trend with \fgas\ \citep[cf.][]{san03,vikhlinin06}

All of the five clusters statistically consistent with having gas fraction
values in excess of the WMAP5 \fbary\ upper bound are passive systems and
all have projected X-ray/BCG offsets exceeding 0.03\rfiveh-- all are also
identified as outliers in Fig.~\ref{fig:compare_r500} (see below). The
cluster gas fraction within \rfiveh\ is typically $\sim$90 per cent of the
cosmic mean \citep[e.g.][]{crain07} and the baryon fraction in massive
clusters is roughly 10 per cent higher than \fgas\
\citep[e.g.][]{lin03,gonzalez07}. Therefore such large gas fractions are
unrealistic and may indicate a failure of the assumption of hydrostatic
equilibrium in the cluster modelling, as might result from significant
dynamical disturbance \citep{nagai07}.  These objects are therefore likely
to be outliers and significant sources of scatter in scaling
relations. Taking the nominal X-ray/BCG offset value of 0.02\rfiveh\ used
to identify the strongest cool core clusters (see
Fig.~\ref{fig:alpha_vs_offset}), the mean gas fraction of the `small
offset' clusters ($\le$\,0.02\rfiveh) is $0.106\pm0.005$ ($\sigma=0.03$)
compared to $0.145\pm0.009$ ($\sigma=0.04$) for the clusters with an offset
$>$\,0.02\rfiveh.

The only emission line BCG cluster with a high gas fraction in
Fig.~\ref{fig:fgas_vs_offset} is Abell~2390 ($\fgas=0.17\pm0.01$, lying
within the WMAP5 \fbary\ band), which is a hot cluster (9.8\,keV) that had
the highest \fgas\ in the 13 cluster sample of
\citet{vikhlinin06} and which is a known merging system, despite hosting a
cool core \citep{allen01c}. As pointed out above, the comparison between
our \fgas\ value and that of \citeauthor{vikhlinin06}
(Fig.~\ref{fig:resid}, right panel) reveals a clear discrepancy within 
\rfiveh. Nevertheless, the agreement is better within \rtwofiveh, and 
our \rfiveh\ value of $1437\pm68$\,kpc is very close to the value of
$1416\pm48$\,kpc calculated by \citeauthor{vikhlinin06}. Furthermore, it
can be seen from inspection of figure~12 in \citet{vikhlinin06} that their
gas fraction profile for Abell~2390 rises sharply just outside \rfiveh,
reaching values consistent with our measurement, albeit at a slightly
larger radius.

To explore the effectiveness of the projected offset between the BCG and
the X-ray centroid in quantifying the cluster dynamical state, we turn to a
comparison between \rfiveh\ determined from two different methods.
Fig.~\ref{fig:compare_r500} shows the comparison between \rfiveh\ as
calculated from the \citeauthor{ascasibar08} cluster model and that
measured by \citet{maughan08} using a scaling relation based on the X-ray
equivalent to the Compton $y$-parameter \citep[\Yx;][]{kravtsov06}. \Yx\ is
simply the total thermal energy of the hot gas within a given radius, and
acts as a robust mass proxy, even for dynamically disturbed clusters
\citep{kravtsov06,poole07,maughan07}. The 30 clusters common to both 
samples are split into four subsamples defined by quartiles of the
projected X-ray centroid/BCG offset. In general, the agreement is
excellent, particular for the clusters in the 50th percentile of X-ray
centroid/BCG projected offset (top panels of Fig.~\ref{fig:compare_r500}),
however six prominent outliers are visible and have been labelled.

The \citeauthor{ascasibar08} cluster model assumes hydrostatic equilibrium
(HSE) to derive \rfiveh, whereas \rfiveh\ inferred from \Yx\ is only
sensitive to the total thermal energy of the gas. A comparison between the
two values therefore reveals the extent to which the assumption of HSE
applies for any given cluster. For five of the outliers in
Fig.~\ref{fig:compare_r500} the \citeauthor{ascasibar08} underpredicts
\rfiveh\ compared to the \Yx-based measurement of \citeauthor{maughan08},
consistent with the tendency for HSE-based measurements to underestimate
the cluster mass \citep{nagai07}. These outliers all fall in the highest
two quartiles of X-ray centroid/BCG projected offset, further supporting
the view that this offset acts as an indicator of dynamical disturbance
\citep[cf.][]{katayama03}. The remaining cluster, Abell~267, has a 
complex mass distribution, based on the gravitational lensing analysis of
\citet{smith05}, which has a highly elliptical morphology. This could
result in the HSE-based X-ray analysis \emph{overestimating} the total
mass, as suggested by the comparison of \rfiveh\ in
Fig.~\ref{fig:compare_r500}, as well as its low \fgas\ ($0.041\pm0.018$;
Fig.~\ref{fig:fgas_vs_offset}), given its large mass
(M$_{500}$=$1.85\pm0.70\times10^{15}$\Msol).

\begin{figure}
\centering
\includegraphics[width=8.7cm]{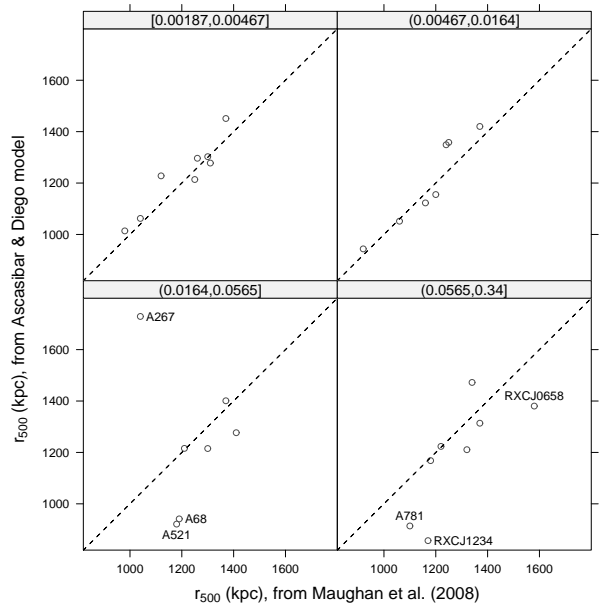}
\caption{ \label{fig:compare_r500}
The comparison between \rfiveh\ calculated from the \citet{ascasibar08}
cluster model and that measured by \citet{maughan08} for the 30 clusters
common to both samples. The clusters have been split into quartiles
according to their projected X-ray centroid/BCG offset, increasing from the
top left to bottom right, as indicated by the range in the strip for each
panel (0.00187--0.00467; 0.00467--0.0164; 0.0164--0.0565; 0.0565--0.34
\rfiveh). The dashed lines indicate the locus of equality.}
\end{figure}

\section{Summary and discussion}
We have shown that the projected X-ray/BCG offset is highly correlated with
the strength of cooling in the host cluster core, as measured by the
logarithmic slope of the gas density profile at 0.04\rfiveh\ ($\alpha$;
Section~\ref{sec:CC}) and that the gas fraction is systematically larger in
clusters with large offsets. In particular, the use of the X-ray 
\emph{centroid} yields a stronger correlation in both these relations, 
compared to using the X-ray \emph{peak} to calculate this offset.

Our results also demonstrate that the activity of BCGs is closely related
to the properties of their host cluster, and specifically the proximity to
and strength of any cool core. The behaviour of the \Halpha\ emitting
galaxies is especially striking in that all 23 of them in our sample of 65
BCGs are found close to ($\la0.02$\rfiveh) a strong ($\alpha\la-0.85$;
Section~\ref{sec:CC}) cool core, with only 1 other non-emission line galaxy
meeting these same criteria.  This is consistent with the work of
\citet{rafferty08} and \citet{cavagnolo08} who find that star formation
only occurs once the hot gas entropy falls below a critical threshold of
$\sim$30\,keV\,cm$^2$. \citet{voit08} show that such a threshold can be
understood if energy output from AGN is coupled to the cooling gas via
thermal conduction, a process which is also capable of thermally
stabilizing \emph{non}-cool core clusters \citep{san09}.

Notwithstanding the effects of conduction or galaxy feedback, it is clear
that star formation is ultimately able to take place when the BCG coincides
closely with a strong cool core. As a consequence, additional enrichment of
the ICM is possible in the vicinity of the BCG, which may explain the
centrally peaked metallicity profiles of cool core clusters
\citep{leccardi08,san09} and the fact that the lowest entropy gas is also 
the most enriched \citep{san09}. The influence of the BCG on the ICM may 
also account for the increased variation in gas properties (e.g. metallicity
and entropy) on small scales \citep[$\sim$0.02\rfiveh;][]{san09}.

The strong connection between line emission and radio emission
(Section~\ref{ssec:BCG_activity}) suggests that star formation and AGN
accretion are fuelled from the same source. However, a significant fraction
(23 per cent; 7/31) of the radio emitting BCGs occupy a narrow range of
relatively large offsets 0.032--0.079\rfiveh\ and have no \Halpha\
emission, suggesting a different origin. These may be cases where recent
merger disruption has triggered a brief burst of AGN
activity.

If the X-ray/BCG projected offset serves to measure the dynamical state of
the cluster, with more disturbed systems having larger values, then its
strong correlation with the steepness of the gas density profile ($\alpha$)
implies that cool core strength progressively diminishes in more
dynamically disrupted clusters. Such a trend would be expected if cluster
mergers were capable of erasing cool cores.  There is a notable dearth in
Fig.~\ref{fig:alpha_vs_offset} of clusters with relatively flat density
profiles and small projected X-ray/BCG offsets. Such cases ought to exist
if AGN outbursts alone were capable of reheating cool cores, or if strong
preheating prevented their formation altogether
\citep{mccarthy08}. On the other hand, if such processes had a disruptive
impact on the core X-ray morphology, then this might shift the centroid as
measured on $\sim$arcmin scales enough to explain the trend.

\section*{Acknowledgments}
AJRS thanks Trevor Ponman and Ria Johnson for useful discussions. We thank
our colleagues in the LoCuSS collaboration for their encouragement and
help, in particular, we thank Eiichi Egami and Victoria Hamilton-Morris for
allowing us to use the CTIO data in advance of their publication, and Chris
Haines for help with BCG identification.  We are grateful to the referee,
Ben Maughan, for a prompt response and helpful suggestions which have
improved the paper. AJRS and GPS acknowledge support from STFC and GPS
acknowledges support from the Royal Society. This work made use of the
NASA/IPAC Extragalactic Database (NED).

\bibliography{/data/ajrs/stuff/latex/ajrs_bibtex}
\label{lastpage}

\end{document}